\documentclass[journal]{IEEEtran}
\ifCLASSINFOpdf
\else
\fi

\newcommand{\normfirst}[1]{\left\|{#1} \right\|_1}

\newcommand{\trace}{\mathsf{trace}}
\usepackage{algorithm2e}
\usepackage{amsthm}

\title{\LARGE \bf
Finite-time Guarantees for Byzantine-Resilient Distributed State Estimation with Noisy Measurements
}

\author{Lili Su and Shahin Shahrampour
\thanks{}
\thanks{Lili Su is with the Computer Science and Artificial Intelligence Laboratory, Massachusetts Institute of Technology,
        Cambridge, MA 02139
        {\tt\small (lilisu@mit.edu).}}%
 \thanks{Shahin Shahrampour is with the Department of Industrial and Systems Engineering,
        Texas A\&M University, College Station, TX 77843
        {\tt\small (shahin@tamu.edu).}}%
}

\usepackage[margin=1in]{geometry}
\usepackage[bookmarks, colorlinks=true, plainpages = false, citecolor = blue,linkcolor=red,urlcolor = blue, filecolor = blue,pagebackref]{hyperref}
\usepackage{url}
\usepackage{amsmath,amsfonts,amssymb,bm, verbatim,dsfont,mathtools}
\usepackage{subfigure}
\usepackage{etoolbox}
\usepackage{tikz}
\usepackage{cite}
\usepackage{xr,xspace}
\usepackage{todonotes}
\usepackage{paralist}
\usepackage{caption,soul}

\newtheorem{theorem}{Theorem}
\newtheorem{lemma}{Lemma}

\newtheorem{definition}{Definition}

\newtheorem{claim}{Claim}

\newtheorem{remark}{Remark}
\newtheorem{assumption}{Assumption}

\usepackage{xspace,prettyref}

\newcommand{\diverge}{\to\infty}

\newcommand{\ones}{\mathbf 1}

\newcommand{\reals}{{\mathbb{R}}}

\newcommand{\expect}[1]{\mathbb{E}\left[ #1 \right]}

\newcommand{\prob}[1]{ \mathbb{P}\left\{ #1 \right\} }

\newcommand{\toas}{\xrightarrow{{\rm a.s.}}}

\newcommand{\iid}{i.i.d.\xspace}
\newrefformat{eq}{(\ref{#1})}
\newrefformat{chap}{Chapter~\ref{#1}}
\newrefformat{sec}{Section~\ref{#1}}
\newrefformat{alg}{Algorithm~\ref{#1}}
\newrefformat{fig}{Fig.~\ref{#1}}
\newrefformat{tab}{Table~\ref{#1}}
\newrefformat{rmk}{Remark~\ref{#1}}
\newrefformat{clm}{Claim~\ref{#1}}
\newrefformat{def}{Definition~\ref{#1}}
\newrefformat{cor}{Corollary~\ref{#1}}
\newrefformat{lmm}{Lemma~\ref{#1}}
\newrefformat{prop}{Proposition~\ref{#1}}
\newrefformat{app}{Appendix~\ref{#1}}
\newrefformat{hyp}{Hypothesis~\ref{#1}}
\newrefformat{thm}{Theorem~\ref{#1}}
\newrefformat{ass}{Assumption~\ref{#1}}

\newcommand{\pth}[1]{\left( #1 \right)}
\newcommand{\qth}[1]{\left[ #1 \right]}
\newcommand{\sth}[1]{\left\{ #1 \right\}}

\newcommand{\abth}[1]{\left | #1 \right |}

\newcommand{\norm}[1]{\left\|{#1} \right\|_2}

\newcommand{\linf}[1]{\left\|{#1} \right\|_{\infty}}

\newcommand{\opnorm}[1]{\left\| #1 \right\|_2}
\newcommand{\iprod}[2]{\left \langle #1, #2 \right\rangle}

\newcommand{\calA}{{\mathcal{A}}}
\newcommand{\calB}{{\mathcal{B}}}

\newcommand{\calE}{{\mathcal{E}}}

\newcommand{\calH}{{\mathcal{H}}}

\newcommand{\calN}{{\mathcal{N}}}

\newcommand{\calR}{{\mathcal{R}}}

\newcommand{\calV}{{\mathcal{V}}}

\renewcommand{\tilde}{\widetilde}

\begin{document}

\maketitle
\thispagestyle{empty}
\pagestyle{empty}

\begin{abstract}
This work considers resilient, cooperative state estimation in unreliable multi-agent networks. A network of agents aims to collaboratively estimate the value of an unknown vector parameter, while an {\em unknown} subset of agents suffer Byzantine faults. Faulty agents malfunction arbitrarily and may send out {\em highly unstructured} messages to other agents in the network. As opposed to fault-free networks, reaching agreement in the presence of Byzantine faults is far from trivial. In this paper, 
we propose a computationally-efficient algorithm that is provably robust to Byzantine faults. 
At each iteration of the algorithm, a good agent (1) performs a gradient descent update based on noisy local measurements, (2) exchanges its update with other agents in its neighborhood, and (3) robustly aggregates the received messages using coordinate-wise trimmed means. Under mild technical assumptions, we establish that good agents learn the true parameter asymptotically in almost sure sense. We further complement our analysis by proving (high probability) {\em finite-time} convergence rate, encapsulating network characteristics. \end{abstract}


\section{Introduction}
Collaborative state/parameter estimation has attracted a considerable attention due to a wide range of applications in internet of things (IoT), wireless networks, power grids, sensor networks, and robotic networks \cite{speranzon2006distributed,xie2012fully,sinopoli2003distributed,olfati2007distributed,kar2012distributed,bullo2009distributed,chen2018internet}. In these applications, a network of (connected) agents collect information in a distributed fashion and share an overarching goal to learn the common {\em unknown} truth $\theta^*\in \reals^d$. Local measurements obtained by each individual agent contain noisy and highly incomplete information about $\theta^*$. 
Nevertheless, the network of agents might be able to collaboratively learn $\theta^*$ by effectively fusing the information contained in their local measurements. 

In the absence of system adversary, the state estimation problem is well-studied \cite{stankovic2011decentralized,kar2012distributed}. However, some practical scenarios such as IoT, micro-grids, and Federated Learning are vulnerable to faults \cite{chen2017distributed}. Motivated by that, we are interested in addressing collaborative estimation in the presence of malicious agents. The existence of malicious agents might arise when some of the networked agents are compromised by a system adversary. Despite the wealth of literature on collaborative estimation with random link failures, packet-dropping failures, and crash failures (e.g. \cite{kar2009distributed}), perhaps less well-known is estimation in the presence of highly {\em unstructured} failures or even {\it adversarial} agents, especially in {\it finite-time} domain.
 
In this work, to formally capture the {\em unstructured} system threat, we adopt Byzantine fault model \cite{Lynch:1996:DA:2821576} -- a canonical fault model in distributed computing. In this model, there exists a system adversary that can choose up to a {\em constant} fraction of agents to compromise and control. An agent suffering Byzantine fault behaves arbitrarily badly by sending out unstructured malicious messages to the good agents. In addition, Byzantine agents may give conflicting messages to different agents in the system. 
Tolerating Byzantine faults is highly non-trivial 
(see e.g. \cite{pease1980reaching,lamport1982byzantine}). For example, it is well-known that in complete graphs, no algorithm can tolerate more than $1/3$ of the agents to be Byzantine \cite{lamport1982byzantine}.
This difficulty arises partially from the system asymmetry caused by the conflicting messages sent by the Byzantine agents. 
In fact, Byzantine consensus with vector multi-dimensional inputs in the complete graphs had not been solved until only recently \cite{mendes2013multidimensional,vaidya2013byzantine}.  

Despite intensive efforts on securing distributed learning (see Section \ref{subsec: related work} for details), to the best of the authors' knowledge, efficient algorithms that are provably resilient to Byzantine faults with less stringent assumptions on {\it noisy} local measurements are still lacking.  In particular, the literature has mostly focused on the asymptotic analysis, leaving the {\it finite-time} guarantees for such algorithms a complementary direction to pursue, which is the main focal point of this work.

\subsection{Our Contributions}
We propose a computationally-efficient algorithm that is provably robust to Byzantine faults. 
At each iteration of our algorithm, a good agent (1) performs a gradient descent update based on local measurements only, (2) exchanges its update with other agents in its neighborhood, and (3) robustly aggregates the received messages using coordinate-wise trimmed means.

For ease of exposition, we first present our results for fully connected networks (complete graphs), and then generalize the obtained results to general networks (incomplete graphs) assuming that the networks satisfy the necessary conditions such that Byzantine-resilient consensus with scalar inputs can be  achievable. 
%
For both cases, we establish that every good agent learns the true parameter asymptotically in the almost sure sense.  
Most importantly, we characterize the {\it finite-time} convergence rate (in high-probability sense), encapsulating network characteristics. 
We finally provide numerical simulations for our method to verify our theoretical results.

\subsection{Related Literature}
\label{subsec: related work}
Resilient estimation, detection, and learning has attracted a great deal of attention in the past few years, and many researchers in the fields of control, signal processing, and network science have addressed the problem by adopting different notions of {\it resilience} or {\it robustness}. 

In \cite{kosut2011malicious,kim2011strategic,sou2013exact}, resilience has been discussed in the context of smart power grid systems using cardinality minimization and its $\ell_1$ relaxations. On the other hand, the focus of \cite{pasqualetti2013attack,fawzi2014secure} is on estimation in Linear Time-Invariant (LTI) systems. In \cite{pasqualetti2013attack}, an interesting approach is proposed for fault detection using monitors, and fundamental monitoring limitations have been characterized using tools from system theory and game theory. 
Furthermore, the approach of \cite{fawzi2014secure} is inspired from the areas of error-correction over the reals and compressed sensing. In \cite{mattingley2010real}, robust Kalman
filtering is discussed, where the estimate updates are derived using a convex $\ell_1$ optimization problem. Authors of \cite{shoukry2016event} consider a model where the observation noise is sparse, in the sense that the faulty sensors have noisy measurements, while other sensors measurements are noiseless. An event triggered projected gradient descent is then proposed to reconstruct the state. In our setting, though the state is fixed, we deal with {\it multi-agent} networks, i.e., the problem must be solved in a {\it distributed} manner since each agent has local (noisy) measurements from the state, and message passing schemes (e.g. consensus) are required to learn the state.

In parallel to advancements on resilient centralized estimation, recent years have witnessed intensive interest in securing distributed estimation. The authors of \cite{sundaram2011distributed} discuss reaching consensus in the presence of malicious agents, assuming a {\it broadcast} model of communication. Chen et al.\ \cite{chen2018resilient} propose a novel adversary detection strategy under which good agents either asymptotically learn the true state or detect the existence of a system fault. If a fault is flagged, the system goes through some external procedure to ``repair'' itself. As a result, the method does not perform estimation under system adversary (which is the focus of this paper). Furthermore, other resilient algorithms have been proposed \cite{chen2018attack,mitra2018resilient,xu2018robust,su2016non,yang2017byrdie} with different assumptions and performance guarantees. Chen et al.\ \cite{chen2018attack} propose an algorithm under which {\em all} of the agents' estimates converge to the true state as long as less than one half of the agents are faulty. However, this algorithm works under the assumption that an agent can {\it fully} observe the true state in the non-faulty condition \cite[Section II.A]{chen2018attack}, as opposed to our model which deals with both observability and noisy measurement issues.  
Mitra and Sundaram \cite{mitra2018resilient} consider the more general LTI systems and characterize the fundamental limits on adversary-resilient algorithms. However, unlike our work, \cite{mitra2018resilient} deals with noiseless observations and the focus is on asymptotic analysis.
Xu et al. \cite{xu2018robust} study the general dynamic optimization problem. They propose a total variation (TV) norm regularization technique to mitigate the effect of malfunctioning agents, but unfortunately, in the static case, the good agents cannot learn the true minimizer (see Corollary 1 in \cite{xu2018robust}). In fact, \cite[Assumption 4]{xu2018robust} might not hold in the sense that under some strategies of the adversary agents, some good agents may appear to be bad to others, and the outgoing links from those agents might be cut off by the good agents. The lack of convergence in this case is consistent with the lower bound result in \cite{su2016fault}.

Another relevant work is the distributed hypothesis testing of \cite{su2016non} where the algorithm {\em Byz-Iter} is proposed. Though this algorithm may work for the state estimation problem, it scales poorly in dimension. Our algorithm is similar to \cite{yang2017byrdie} in that we both combine local gradient descent with coordinate-wise message trimming. Although \cite{yang2017byrdie} considers a more general optimization framework, it is implicitly assumed that the optimization problem can be separated into independent optimization problems (of the size of unknown parameter); otherwise, \cite[Lemma 1]{yang2017byrdie} does not hold and the proof in \cite{su2016fault} cannot be applied.

\section{Problem Formulation}

\subsection{Network Model}
We consider a multi-agent network which is a collection of $n$ agents/nodes communicating with each other through a communication network 
$G(\calV, \calE)$, where $\calV = \{1, \cdots, n\}$ and $\calE$ denote the set of nodes and edges, respectively. We denote by $\calN_i$
the set of incoming neighbors of agent $i$. 
An unknown subset of agents of size at most $b$, denoted by $\calA$, might be 
{\it bad} or {\it adversarial}.  The set 
$\calA$ is chosen by the system adversary. 
For ease of exposition, let 
\[
|\calV/\calA| = \phi. 
\]
Clearly, $\phi \ge n -b$. 

Good agents (agents in $\calV/\calA$) aim to estimate the unknown parameter collaboratively, but bad agents (agents in $\calA$) 
can adversarially affect the estimation procedure by sending {\it completely arbitrary}, {\it malicious}, and possibly {\it conflicting} messages to the good agents.

\subsection{Observation Model}
In this work, we focus on a linear observation model, where $y_i(t)$ represents the local measurement of agent $i$ at time $t$ as follows
\begin{align}
\label{eq: linear observation noiseless}
y_i(t) = H_i \theta^* + w_i(t),
\end{align}
and $H_i \in \reals^{n_i\times d}$ is the local observation matrix. The noise sequence $w_i(t)$ is $\iid$ with $\expect{w_i(t)} = {\bf 0}$ and $\expect{w_i(t)w_i(t)^\top} = \Sigma_i$. The sequences are bounded for all agents, i.e., there exists constant $C>0$ such that $\prob{\norm{w_i(t)}\le C} =1$ for $i\in \calV$. Moreover, the noise sequences across 
good agents are independent. That is, $\pth{w_i(t), t\ge 1}$ and $\pth{w_j(t), t\ge 1}$ for $i\not= j$ are independent. As in practice the observation matrix $H_i$ is often fat, i.e., $n_i \ll d$, each agent $i$ must obtain information from others 
to correctly estimate $\theta^*$.

\subsection{Fault Model}\label{subsec:threat model}
To formally capture the system threat, we adopt the Byzantine fault model \cite{Lynch:1996:DA:2821576} -- a canonical fault model in distributed computing. 
In this model, there exists a system adversary that can choose up to $b$ of the $n$ agents (where $b< n$) to compromise and control. 
Recall that this set of agents is denoted by $\calA$. 
An agent suffering Byzantine fault is referred to as Byzantine agent. 
While the set $\calA$ is unknown to good agents, a standard assumption in the literature is that the value of $b$ is common knowledge \cite{Lynch:1996:DA:2821576}.

The system adversary is extremely {\it powerful} in the sense that it has complete knowledge of the network, including the local program that each good agent is supposed to run, the true value of the parameter $\theta^*$, the current status and running history of the multi-agent network system, the running history, etc. Hence, the Byzantine agents can collude with each other and deviate from their pre-specified local programs to {\em arbitrarily} misrepresent information to the good agents. In particular, Byzantine agents can mislead each of the good agents in a unique fashion, i.e., letting $m_{ij}(t) \in \reals^d$ be the message sent from agent $i \in \calA$ to agent $j \in \calV \setminus \calA$ at time $t$, it is possible that $m_{ij}(t)\not= m_{i{j^{\prime}}}(t)$ for $j\not=j^{\prime} \in  \calV \setminus \calA$. 
\begin{remark} 
Note that due to the {\it extreme freedom} given to Byzantine agents and the system asymmetry caused by them, a resilient distributed  
solution to the estimation problem is highly non-trivial even in complete graphs. 
In particular, it is well-known that in complete graphs, no algorithm can tolerate more than $1/3$ of the agents to be Byzantine \cite{lamport1982byzantine}. 
\end{remark}

\subsection{Finite-time vs. Asymptotic Local Functions} 
The Byzantine-resilient state estimation problem can be viewed with an optimization lens, where each good agent would only asymptotically know its local function. For each agent $i\in \calV $, define the {\it asymptotic} local function $f_i: \reals^d \to \reals$ as
\begin{align}
\label{eq: local cost}
f_i(x) & \triangleq \expect{\frac{1}{2}\norm{H_i x - y_i}^2}, 
\end{align}
where the expectation is taken over the randomness of $w_i$. Note that $f_i$ is well-defined for each $i\in \calV$ regardless of whether it is suffering Byzantine faults or not. 
Since the distribution of $w_i$ is unknown to agent $i$, at any finite $t$, function $f_i$ is not accessible to agent $i$. 
However, the agent has access to the  {\it finite-time} or {\it empirical} local function 
\begin{align}
\label{eq: empirical}
f_{i,t} (x) \triangleq  \frac{1}{t} \sum_{s=1}^t \frac{1}{2} \norm{H_i x - y_i(s)}^2, 
\end{align}
whose gradient at $x$ is 
\begin{align}
\label{eq: empirical gradient}
\nabla f_{i,t} (x) 
& = H_i^{\top}H_i \pth{x - \theta^*} - H_i^{\top}\frac{1}{t}\sum_{r=1}^t w_i(t).
\end{align}

\section{Byzantine-Resilient State Estimation}

To robustify distributed state estimation against Byzantine faults, one approach may be to combine the local gradient descent with multi-dimensional Byzantine-resilient consensus \cite{su2016non, vaidya2013byzantine, mendes2013multidimensional} (which typically relies on using Tverberg points). 
However, the performance of any such algorithm is proved to scale poorly in the dimension of the parameter $d$ \cite{su2016non, vaidya2013byzantine, mendes2013multidimensional}.  
This is partially due to the fact that 
different dimensions of the inputs strongly interfere with each other, and the Byzantine agents can inject wrong information with both extreme magnitudes and directions.

To improve the scalability with respect to $d$ and to improve the computation complexity, instead of multi-dimensional Byzantine-resilient consensus, we robustly aggregate the received messages using coordinate-wise trimmed means.  

\subsection{Algorithm}

We propose an algorithm, named {\em Byzantine-resilient state estimation}, where each good agent iteratively aggregates the received messages. 
To robustify, the agent discards the largest $b$ and the smallest $b$ values for each component. In particular, in each iteration, an agent performs the following three steps: 
\begin{itemize}
\item  {\em Local gradient descent: } Agent $i$ first computes the noisy local gradient $\nabla f_{i,t} (x_i(t-1))$, and performs local gradient descent to obtain $z_i(t)$, i.e., 
\begin{align*}
z_i(t) = x_i(t-1) - \nabla f_{i,t} (x_i(t-1)). 
\end{align*}
Note that the step-size used in this update is 1.  
\item  {\em Information exchange:}  It exchanges $z_i(t)$ with other agents in its local neighborhood. Recall that $m_{ij}(t) \in \reals^d$ is the message sent from agent $i$ to agent $j$ at time $t$. It relates to $z_i(t)$ as follows: 
\begin{align*}
m_{ij}(t) = 
\begin{cases}
z_i(t) & ~~~ \text{if} ~ i\in (\calV/\calA)  ; \\
\star & ~~~ \text{if} ~ i\in \calA, 
\end{cases}
\end{align*}
where $\star$ denotes an arbitrary value. Byzantine agents can mislead good agents differently, i.e., if $i\in \calA$, it might hold that $m_{ij}(t)\not= m_{i{j^{\prime}}}(t)$ for $j\not=j^{\prime}\in \calV\setminus\calA$.

\item  {\em Robust aggregation:} For each component $k=1, \ldots, d$, the agent computes the trimmed mean and uses them to obtain $x_i(t)$. 
\end{itemize}
The formal description of the algorithm for agent $i\in \calV\setminus\calA$ is given in Algorithm \ref{alg: Consensus-based algorithm}. 

\RestyleAlgo{boxruled}

\begin{algorithm}[ht]
\caption{Byzantine-resilient state estimation} 

\label{alg: Consensus-based algorithm}
\noindent {\bf Input:} $b$ and $T$ \\
\noindent {\bf Initialization:} Set $x_i(0)$ to an arbitrary value for each agent $i\in \calV$ 

\For{$t=1,\ldots,T$}{

- Obtain a new measurement $y_i(t)$\; 

- Compute the local noisy gradient $\nabla f_{i,t} (x_i(t-1))$ according to \eqref{eq: empirical gradient}\; 

- Compute $z_i(t) =x_i(t-1) - \nabla f_{i,t} (x_i(t-1))$\; 

- Send $z_i(t)$ to its outgoing neighbors\; 

\For{$k=1, \ldots, d$}
{

- Sort the $k$--th component of the received messages $m_{ji}(t)$ for $j\in \calN_i \cup\{i\}$  
in a non-decreasing (increasing) order\; 
- Remove the largest $b$ values and the smallest $b$ values\; 

- Denote the remained ``agent'' indices set as $\calR_i^k(t)$ and set  $$x_i^k(t) = \frac{1}{\abth{\calR_i^k(t)}} \sum_{j\in \calR_i^k(t)} \iprod{m_{ji}(t)}{e_k}.$$  
}
- Set $(x_i(t))^{\top} = \pth{x_i^1(t), \ldots, x_i^d(t)}$. }
\noindent {\bf Output:} $x_i(T)$. 
\end{algorithm}

\section{Finite-time Guarantee for Complete Networks}
In this section, we provide results for the case that $G(\calV, \calE)$ is a complete graph. Beside the fact that the technical analysis of complete graphs would be different from that of incomplete graphs (in terms of assumptions), the former is particularity interesting in computer networks. In fact, in many computer networks efficient communication protocols (such as TCP/IP) can be implemented such that any two computer are logically connected.

It can be shown that the update of $x_i$ uses the information provided by the {\em good} agents only. In addition, each of the good agent has limited impact on $x_i$, formally stated next. 
\begin{lemma}
\label{lm: weight upper bound}
For each iteration $t$, each good agent $i\in \calV/\calA$, and each $k$, 
there exist coefficients $\pth{\beta_{ij}^k(t), ~ j\in \calV/\calA}$ such that 
\begin{itemize}
\item $x_i^k(t) = \sum_{j \in \calV/\calA} \beta_{ij}^k(t)  \iprod{z_j(t)}{e_k}$; 
\item $0\le \beta_{ij}^k(t)  \le \frac{1}{\phi -b}$ for all $j \in \calV/\calA$  and $\sum_{ j\in \calV/\calA }\beta_{ij}^k(t) =1$. 
\end{itemize}
\end{lemma}
Notice that the sets of convex coefficients for different values of $k$ might be different, i.e., 
$\pth{\beta_{ij}^k(t), ~ j\in \calV/\calA} \not= \pth{\beta_{ij}^{k^{\prime}}(t), ~ j\in \calV/\calA}$ for $k\not=k^{\prime}$. Moreover, even for the same $k$, the convex coefficients might be different for different good agents, i.e., $\pth{\beta_{ij}^k(t), ~ j\in \calV/\calA} \not= \pth{\beta_{i^{\prime}j}^{k}(t), ~ j\in \calV/\calA}$ for $i\not=i^{\prime}$. This stems from the freedom of Byzantine agents in sending  different messages across agents, i.e., $m_{aj}\not=m_{aj^{\prime}}$ if $a\in \calA$ and $j\not=j^{\prime}$.

To prove the convergence of Algorithm \ref{alg: Consensus-based algorithm}, we use the following assumption. 
\begin{assumption}
\label{ass: alternative assumption}
For all $k=1, \cdots, d$, we have that
\begin{align*}
 \frac{1}{\phi -b}   \sum_{j\in \calV/\calA} \normfirst{\pth{{\bf I} - H_j^{\top} H_j}e_{k} } <1.  
 \end{align*}
\end{assumption}

Note that $\normfirst{\pth{{\bf I} - H_j^{\top} H_j}e_{k} }$ is the $\ell_1$ norm of the $k$--th column of matrix ${\bf I} - H_j^{\top} H_j$.  It can well be the case that $\normfirst{\pth{{\bf I} - H_j^{\top} H_j}e_{k} } \geq 1$ for some good agents. 
However, Assumption \ref{ass: alternative assumption} implies that for each $k=1, \cdots, d$, there exists at least $b+1$ good agents such that 
\begin{align*}
\normfirst{\pth{{\bf I} - H_j^{\top} H_j}e_{k}} <1. 
\end{align*}
The above assumption is imposed for the $d$ components individually. None of the agents are required to satisfy $\normfirst{\pth{{\bf I} - H_j^{\top} H_j}e_{k}} <1$ simultaneously for all $k=1, \cdots, d$. Now, let 
\begin{align}
\label{eq: rate def}
\rho ~ \triangleq ~  \max_{k: 1\le k \le d}\frac{\sum_{j\in \calV/\calA} \normfirst{\pth{{\bf I} - H_j^{\top} H_j}e_{k}}}{\phi -b}. 
\end{align}
Clearly, $\rho<1$ under Assumption \ref{ass: alternative assumption}. 
For ease of exposition, for each $j\in \calV/\calA$ and for any $\lambda \in (0,1)$, let 
\begin{align}
\label{eq: cumulative noise}
R_j(\lambda, t) ~ \triangleq ~ \sum_{m=0}^{t-1} \lambda^m \norm{\frac{\sum_{r=1}^{t-m} w_j(r)}{t-m}}. 
\end{align}
The following two concentration results are two key auxiliary lemmas for our main theorem. 
\begin{lemma}
\label{lm: noise bound}
Suppose Assumption \ref{ass: alternative assumption} holds. Then, for each $j\in \calV/\calA$ and for any $\lambda \in (0,1)$
\begin{align*}
\lim_{t\diverge} R_j(\lambda, t) ~ = ~ 0 ~~ \text{almost surely}. 
\end{align*} 
\end{lemma}
In addition, we characterize the {\it finite-time} convergence rate of $R_j(\lambda, \cdot)$ for any fixed $\lambda$.  
\begin{lemma}
\label{lm: convergence rate}
Suppose Assumption \ref{ass: alternative assumption} holds. Then for each $j\in \calV/\calA$  and for any $\lambda \in (0,1)$
\begin{align*}
&\prob{R_j(\lambda, t) \ge \sqrt{\trace(\Sigma_j)}\sum_{m=1}^{t-1} \lambda^m \frac{1}{\sqrt{t-m}} + \epsilon} \\
 &\le \exp \pth{\frac{-\epsilon^2(1-\lambda)^2t}{8C^2}},
\end{align*}

\end{lemma}
Lemma \ref{lm: convergence rate} implies that $\forall j\in \calV/\calA$, with probability at least $1-\delta$,  $R_j(t) = O\pth{\sqrt{\pth{\log \frac{1}{\delta}}/t}}$.

\begin{theorem}
\label{thm: complete consensus alt}
Suppose Assumption \ref{ass: alternative assumption} holds and the graph $G(\calV, \calE)$ is complete. Then
$$\max_{i\in \calV/\calA} \linf{x_i(t) - \theta^*}  \toas 0.$$
Moreover, with probability at least \\
$1-\phi\exp \pth{\frac{-\epsilon^2(1-\rho)^2t}{8C^2}}$, it holds that 
\begin{align*}
&\max_{i\in \calV/\calA} \linf{x_i(t) - \theta^*} 
 \le  \rho^{t} \max_{i\in \calV/\calA} \linf{x_i(0) - \theta^*} \\
 &\quad + C_0\pth{\sum_{i\in \calV/\calA} \sqrt{\trace(\Sigma_j)}} \sum_{m=1}^{t-1} \frac{\rho^m}{\sqrt{t-m}} + \phi\epsilon, 
\end{align*}
where 
$C_0 ~ \triangleq ~ \max_{i\in \calV/\calA} \opnorm{H_i}$. 
\end{theorem}
The theorem indicates that all good agents (in a complete graph) are eventually able to learn the true parameter $\theta^*$ almost surely. Also, with high probability the rate can be characterized as above, providing a {\it finite-time} guarantee for resilient estimation. The finite-time bound captures the performance, in terms of $\Sigma_j$, the noise covariance for agent $j \in \calV/\calA$, as well as $\rho$, which can crudely serve as a measure of observability in view of \eqref{eq: rate def}.

\section{Finite-time Guarantees for Incomplete Networks}
\label{sec: general graphs}

\subsection{Incomplete Graphs: Multihop Communication}
So far, our analysis of Algorithm \ref{alg: Consensus-based algorithm} has focused on complete graphs. 

For computer networks, this is a reasonable assumption as computers are connected to each other through some communication (routing) protocols. 
Our results are also applicable to wireless networks under some implementation assumptions. 

Concretely, let  $G(\calV, \calE)$ be the physical network that is not fully connected. 
Suppose the networked agents are allowed to relay the messages sent by others such that multi-hop communication can be implemented. We can adopt coding to force the Byzantine agents to either refuse to relay information or faithfully relay the messages without alternation \cite{pease1980reaching}. Thus, as long as the node-connectively of $G(\calV, \calE)$ is at least $b+1$, each good agent can reliably receive messages from other good agents in the network. We can use our algorithm to robust aggregate the received messages and perform one-step update. Similar analysis applies. 

\subsection{Incomplete Graphs: Local Communication}
Message forwarding might be costly or even infeasible for some wireless networks. Algorithms that rely solely on local communication are still highly desirable.
Fortunately, with reasonable assumptions, Algorithm \ref{alg: Consensus-based algorithm} works. 
Our algorithm is a consensus-based algorithm, so to make the paper self-contained, we briefly review relevant existing results on Byzantine consensus. 

\subsubsection{Byzantine Consensus with Scalar Inputs}
Note that, in contrast to fault-free consensus, Byzantine-resilient consensus with scalar inputs and with multidimensional inputs are fundamentally different \cite{mendes2013multidimensional,vaidya2013byzantine,vaidya2012iterative}. Our algorithm relies on Byzantine-resilient consensus with scalar inputs.

Tight topological conditions are characterized in \cite{vaidya2012iterative}, where the conditions are stated in terms of a family of subgraphs of $G(\calV, \calE)$. Those subgraphs capture the ``real'' information flow under the message trimming strategy. Informally speaking, trimming certain messages can be viewed as ignoring (or removing) incoming links that carry the outliers. The non-uniqueness of the subgraph arises partially from the fact that the Byzantine agents can behave adaptively and arbitrarily. 
Such subgraphs are referred to as {\em reduced graphs}, defined as follows.  
\begin{definition}\cite{vaidya2012iterative}
\label{reduced graph}
A reduced graph $H$ of $G(\calV, \calE)$ is obtained by (i) removing all faulty nodes $\calA$, and all the links incident on the faulty nodes $\calA$; and (ii) for each non-faulty node (nodes in $\calV/\calA$), removing  up to $b$ additional incoming links.
\end{definition}
It is important to note that the non-faulty agents {\it do not} know the identities of the faulty agents. 
Let $\calH$ be the collection of all reduced graphs of $G(\calV, \calE)$, and let 
\[
\xi:= \abth{\calH}. 
\]
\begin{definition}
\label{source}
A source component in a given reduced graph is a strongly connected component, 
which does not have any incoming links from outside of that component.
\end{definition}
It turns out that the effective communication network is potentially time-varying (partly) due to time-varying
behaviors of Byzantine agents. 
The tight network topology condition for scaler valued consensus to be achievable is characterized in \cite{vaidya2012iterative}.   
\begin{theorem} \cite{vaidya2012iterative}
\label{thm: iabc}
For scalar inputs, iterative approximate Byzantine consensus is achievable among non-faulty agents if and only if every reduced graph of $G(\calV, \calE)$ contains only one source component.
\end{theorem}
Under the condition in Theorem \ref{thm: iabc}, it follows that in any reduced graph, a node in the source component can reach every other nodes. 

\subsubsection{Correctness of Algorithm \ref{alg: Consensus-based algorithm} for Incomplete Graphs}

We will show the correctness of our Algorithm \ref{alg: Consensus-based algorithm} assuming that Byzantine consensus with scalar inputs is achievable over $G(\calV, \calE)$, and the following assumption holds. 
\begin{assumption}
\label{ass: general graphs}
For each non-faulty node $j\in \calV/\calA$ and each $k=1, \cdots, d$, 
\begin{align*}
\normfirst{\pth{{\bf I} -H_j^{\top} H_j} e_{k}} ~ \le ~ 1. 
\end{align*}
In addition,  any reduced graph $H$ contains a node in its unique source component such that for all $k=1, \cdots, d$, 
\begin{align*}
\abth{(\calN_i\cup \{i\} /\calA)\cap \sth{j:\normfirst{\pth{{\bf I} -H_j^{\top} H_j}e_{k}} <1}}
 \ge b+1.  
\end{align*}
\end{assumption}
Note that in Assumption \ref{ass: general graphs}, $\calN_i$ is the incoming neighbors of node $i$ in the original graph $G(\calV, \calE)$. 

Define $\rho_0$ as 
\begin{align}
\label{eq: rate general}
\rho_0 := \max_{1\le k \le d} \max_{j: \normfirst{\pth{{\bf I} -H_j^{\top} H_j} e_{k}} <1} \normfirst{\pth{{\bf I} -H_j^{\top} H_j} e_{k}}.  
\end{align}
In \eqref{eq: rate general}, the maximization 
\[\max_{j: \normfirst{\pth{{\bf I} -H_j^{\top} H_j} e_{k}} <1} \normfirst{\pth{{\bf I} -H_j^{\top} H_j} e_{k}}
\]
is taken over the non-faulty nodes only. 

Similar to the analysis for the complete graphs, it can be shown that the update of $x_i$ uses the information provided by its good neighbors only. 
\begin{lemma}\cite[Claim 2]{vaidya2012matrix}
\label{lm: weight upper bound general}
For each iteration $t$, each good agent $i\in \calV/\calA$, and each $k$, 
there exist coefficients $\pth{\beta_{ij}^k(t), ~ j\in \calN_i \cup\{i\}}$ such that 
\begin{itemize}
\item $x_i^k(t) = \sum_{j \in \calN_i \cup\{i\}/\calA} \beta_{ij}^k(t)  \iprod{z_j(t)}{e_k}$; 
\item There exists a subset of $\calB_i(t) \subseteq \calN_i \cup\{i\}/\calA$ such that $|\calB(t)| \ge |\calN_i \cup\{i\}/\calA| -b$ and $ \beta_{ij}^k(t) \ge \frac{1}{2(|\calN_i \cup\{i\}/\calA| -b) }$ for each $j\in \calB_i(t)$. 
\end{itemize}
\end{lemma}

In the next theorem, we establish that (under the assumption above) the estimates of all agents are consistent almost surely, and furthermore, we characterize the (high probability) {\it finite-time} convergence rate of these estimates.  
\begin{theorem}
\label{thm: general graphs local communication}
Suppose that every reduced graph of $G(\calV, \calE)$ contains a single source component, and Assumption \ref{ass: general graphs} holds. 
Then
$$\max_{i\in \calV/\calA} \linf{x_i(t) - \theta^*}  \toas 0.$$
Let $\gamma : = 1- \frac{1-\rho_0}{\pth{2(\phi-b)}^{\xi\phi}}$. 
With probability at least \\
$1-\phi\exp \pth{\frac{-\epsilon^2(1-\gamma^{\frac{1}{\xi\phi}})^2t}{8C^2}}$, it holds that 
\begin{align*}
&\max_{i\in \calV/\calA} \linf{x_i(t) - \theta^*} 
 \le  \gamma^{\frac{t}{\xi\phi}} \max_{i\in \calV/\calA} \linf{x_i(0) - \theta^*} \\
 &\quad + C_0\pth{\sum_{i\in \calV/\calA} \sqrt{\trace(\Sigma_j)}} \sum_{m=1}^{t-1} \frac{\gamma^{\frac{m}{\xi\phi}}}{\sqrt{t-m}} + \phi\epsilon. 
\end{align*}
\end{theorem}

\section{Numerical Experiments}
We now provide empirical evidence in support of our algorithm. We consider a complete graph of $|\calV\setminus\calA|=30$ agents. Each component of the unknown parameter $\theta^*\in \reals^{50}$ is generated randomly within the interval $[-1,1]$ and is fixed thereafter during the estimation process. Moreover, the observation matrices $H_i\in \reals^{20\times 50}$ for each $i$ are chosen such that Assumption \ref{ass: alternative assumption} holds.  

Evidently, in this example, $n_i=20$ for all $i$. Throughout, the adversarial agents can send out completely arbitrary messages in lieu of true gradients. We generate these arbitrary messages using a random $50$-dimensional vector, each component of which is sampled from $\calN(0,9)$.  

\begin{figure}[t!]
\begin{center}
        \includegraphics[trim={12mm 0mm 0 0},clip,scale=.27]{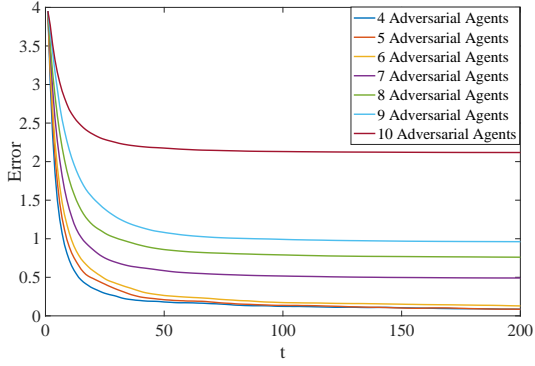}
\end{center}
\vspace{-0.3cm}
	\caption{The plot of error decay versus time for different number of adversarial agents.}
	\label{plot}
\end{figure}

Let us now define the network performance metric as 
$$
\text{Error}(t)\triangleq \frac{1}{\phi}\sum_{i\in \calV\setminus\calA} \|\theta^*-x_i(t)\|
$$
and plot in Fig. \ref{plot} the error for various values of adversarial agents $|\calA|\in \{4,5,\ldots,10\}$. We observe a dichotomy, where for $|\calA|<7$ the error converges to zero, whereas for $|\calA|\geq 7$ the convergence does not occur. Moreover, increase in the number of adversarial agents degrades the performance.

\section{Conclusion}
We studied resilient distributed estimation, where a network of agents want to learn the value of an unknown parameter in the presence of Byzantine agents. The main challenges in the problem are as follows: (i) Byzantine agents send out arbitrary messages to other agents, (ii) good agents need to deal with noisy measurements, and (iii) the parameter is not locally observable. We proposed an algorithm that allows agents to collectively learn the true parameter asymptotically in almost sure sense, and we further complemented our results with {\it finite-time} analysis. Future directions include resilient estimation and learning in a more general setting, where agents observations can be a nonlinear function of the unknown parameter.    
Another interesting direction is to investigate the minimal condition needed on the local observation matrices of the good agents for the problem to be solvable. 

\appendix

\section*{Proof of Lemma \ref{lm: weight upper bound}}
\label{app: proof of lemma - weight upper bound}
We prove this lemma by construction. Note that this construction is only used in the algorithm analysis rather than an algorithm input. That is, to run the algorithm, each agent (either good or faulty) does not need to know $\beta$. 


For ease of exposition, let $[\calR_i^k(t)]^+$ and  $[\calR_i^k(t)]^-$ be the non-overlapping subsets of $\calV$ whose gradient's $k$--th entry 
are trimmed away by agent $i$. 
Precisely, \\
(a) $\abth{[\calR_i^k(t)]^-} = b =\abth{[\calR_i^k(t)]^+}$;   \\
(b) $[\calR_i^k(t)]^-$, $[\calR_i^k(t)]^+$ and $\calR_i^k(t)$ partition set $\calV$; \\
(c) $\forall$ $j^{\prime} \in [\calR_i^k(t)]^-, j\in \calR_i^k(t), ~ \text{and }  j^{\prime \prime} \in [\calR_i^k(t)]^+$ it holds that 
\begin{align}
\label{eq: ordering}
\iprod{m_{j^{\prime}i}(t)}{e_k} \le \iprod{m_{ji}(t)}{e_k} \le \iprod{m_{j^{\prime \prime}i}(t)}{e_k}.
\end{align} 

\vskip 0.6\baselineskip 

We consider two cases: (1) $\calR_i^k(t) \cap \calA = \emptyset$; and (2) $\calR_i^k(t) \cap \calA  \not= \emptyset$. 

\noindent{\bf Case 1:} Suppose that $\calR_i^k(t) \cap \calA = \emptyset$. 
We construct the convex coefficients as follows: 

\noindent{\bf Case 1-1:}
When $\abth{\calA} =b$, we have $\phi -b =  n-2b.$
We choose the convex coefficients as 
\begin{align*}
\beta_{ij}^k(t)   = 
\begin{cases}
& \frac{1}{n-2b},  ~~\forall j\in  \calR_i^k(t), \text{and}\\
& 0, ~~ \forall j\notin  \calR_i^k(t). 
\end{cases}
\end{align*}
Clearly, in this construction, 
$\beta_{ij}^k(t) \le \frac{1}{\phi -b}$. 

\noindent{\bf Case 1-2:}
When $\abth{\calA} < b$, 
it holds that  
\begin{align}
\label{eq: trimmed nonfaulty 1}
\abth{[\calR_i^k(t)]^- / \calA} \ge b- |\calA|, 
\end{align}
and 
\begin{align}
\label{eq: trimmed nonfaulty 1}
\abth{[\calR_i^k(t)]^+ / \calA} \ge b- |\calA|. 
\end{align}
By \eqref{eq: ordering}, we have   
\begin{align*}
&\frac{1}{\abth{[\calR_i^k(t)]^- /\calA} } \sum_{j\in [\calR_i^k(t)]^- /\calA} \iprod{z_j(t)}{e_k} \\
&\le  \frac{1}{n-2b} \sum_{j\in \calR_i^k(t)}\iprod{z_j(t)}{e_k} \\
&\le \frac{1}{\abth{[\calR_i^k(t)]^+ /\calA} } \sum_{j\in [\calR_i^k(t)]^+ /\calA} \iprod{z_j(t)}{e_k}.    
\end{align*}
Thus, there exists $\alpha \in [0, 1]$ such that 
\begin{align}
\label{eq: rewriting}
\nonumber
 &\frac{1}{n-2b} \sum_{j\in \calR_i^k(t)}\iprod{z_j(t)}{e_k}  \\
 \nonumber
 & =  \frac{\alpha}{\abth{[\calR_i^k(t)]^- /\calA} } \sum_{j\in [\calR_i^k(t)]^- /\calA} \iprod{z_j(t)}{e_k} \\
 & +  \frac{1-\alpha}{\abth{[\calR_i^k(t)]^+ /\calA} } \sum_{j\in [\calR_i^k(t)]^+ /\calA} \iprod{z_j(t)}{e_k}. 
\end{align}

Note that 
\begin{align*}
&\frac{1}{n-2b} \sum_{j\in \calR_i^k(t)}\iprod{z_j(t)}{e_k}  \\
& = \frac{1}{\phi-b} \pth{1+ \frac{f-|\calA|}{n-2b}}  \sum_{j\in \calR_i^k(t)}\iprod{z_j(t)}{e_k}\\
& = \frac{1}{\phi-b} \sum_{j\in \calR_i^k(t)}\iprod{z_j(t)}{e_k} \\
& +  \frac{1}{\phi-b}\frac{b-|\calA|}{n-2b}   \sum_{j\in \calR_i^k(t)}\iprod{z_j(t)}{e_k}\\
& \overset{(a)}{=}  \frac{1}{\phi-b} \sum_{j\in \calR_i^k(t)}\iprod{z_j(t)}{e_k} \\
& + \frac{\alpha(b-|\calA|)}{(\phi-b)\abth{[\calR_i^k(t)]^- /\calA} }
 \sum_{j\in [\calR_i^k(t)]^- /\calA} \iprod{z_j(t)}{e_k} \\
 &+ \frac{(1-\alpha)(b-|\calA|)}{(\phi-b)\abth{[\calR_i^k(t)]^+ /\calA}} \sum_{j\in [\calR_i^k(t)]^+ /\calA} \iprod{z_j(t)}{e_k}. 
\end{align*}
where equality (a) follows from \eqref{eq: rewriting}. Choose the convex coefficients for the good agents as follows: 
\begin{align*}
\beta_{ij}^k(t)   = 
\begin{cases}
& \frac{1}{\phi-b},  ~~\forall j\in  \calR_i^k(t), \\
& \frac{\alpha(b-|\calA|)}{(\phi-b)\abth{[\calR_i^k(t)]^- /\calA} } ~~\forall j\in  [\calR_i^k(t)]^- /\calA, \\
& \frac{(1-\alpha)(b-|\calA|)}{(\phi-b)\abth{[\calR_i^k(t)]^+ /\calA} } ~~\forall j\in  [\calR_i^k(t)]^+ /\calA. 
\end{cases}
\end{align*}
The fact that $\alpha$ is unknown does not affect our correctness proof -- as our algorithm not use these coefficients as input. We use the existence of $\alpha$ for analysis. 
It is easy to see that the above coefficients are valid convex coefficients. It remains to check that $\beta_{ij}^k(t)\le \frac{1}{\phi-b}$ for all $j\in \calV /\calA$. For all good in $\calR_i^k(t)$, clearly $\beta_{ij}^k(t) \le \frac{1}{\phi-b}$. 
For $j\in  [\calR_i^k(t)]^- /\calA$, by \eqref{eq: trimmed nonfaulty 1} and the fact that $\alpha \le 1$, we have 
\begin{align*}
\beta_{ij}^k(t)  
\le \frac{\alpha(b-|\calA|)}{(\phi-b)(b-|\calA|)}  
\le \frac{1}{\phi-b}, 
\end{align*}
Similarly, we can show  $\beta_{ij}^k(t)  \le  \frac{1}{\phi-b}$ for $j\in  [\calR_i^k(t)]^+ /\calA$.

Case 2 can be proved similarly.

\section*{Proof of Lemma \ref{lm: noise bound}}
Let $\omega$ be any sample path such that $\lim_{t\diverge} \frac{1}{t} \sum_{r=1}^t w_j(r, \omega) = 0$. Note that fixing $\omega$, $w_j(t,\omega)$ for $t=1, \cdots$ is a standard sequence of vectors. We will show that 
\begin{align}
\label{eq: path convergence}
\lim_{t\diverge}\sum_{m=0}^{t-1} \lambda^m \norm{\frac{\sum_{r=1}^{t-m}w_j(r, \omega)}{t-m}} ~ = ~ 0.
\end{align}
By Strong Law of Large Number we know that 
\begin{align*}
\prob{\omega\in \Omega:  \lim_{t\diverge} \frac{1}{t} \sum_{r=1}^t w_j(r, \omega) = 0 }  =  1. 
\end{align*}
Thus, if \eqref{eq: path convergence} holds, then 
\begin{align*}
\prob{\omega\in \Omega:  \lim_{t\diverge}\sum_{m=0}^{t-1} \lambda^m \norm{\frac{\sum_{r=1}^{t-m}w_j(r, \omega)}{t-m}} = 0 } = 1,
\end{align*}
proving the lemma. 

Next we show \eqref{eq: path convergence}. It is enough to show that for any $\epsilon >0$, there exists $t\ge t(\epsilon, \omega)$ such that 
\begin{align}
\label{eq: epsilon convergence}
\sum_{m=0}^{t-1} \lambda^m \norm{\frac{1}{t-m}\sum_{r=1}^{t-m} w_j(r)} \le \epsilon. 
\end{align}
Since $\lim_{t\diverge} \frac{1}{t} \sum_{r=1}^t w_j(r, \omega) = 0$, for any $\frac{(1-\lambda)\epsilon}{2}$, there exists $t_0(\epsilon, \omega)$ such that for any $t\ge t_0(\epsilon, \omega)$, 
\begin{align*}
\norm{\frac{1}{t}\sum_{r=1}^{t} w_j(r)} \le  \frac{(1-\lambda)\epsilon}{2}. 
\end{align*}
In addition, for any $t\ge t_0(\epsilon, \omega)$, it holds that 
\begin{align*}
&\sum_{m=0}^{t-1} \lambda^m \norm{\frac{\sum_{r=1}^{t-m} w_j(r)}{t-m}} \\
& \le \sum_{m=0}^{t-t_0(\epsilon, \omega)} \lambda^m \frac{(1-\lambda)\epsilon}{2} + C \sum_{m=t-t_0(\epsilon, \omega) + 1}^{t-1} \lambda^m\\
& \le \frac{\epsilon}{2} +  C \frac{\lambda^{t-t_0(\epsilon, \omega) + 1}}{1-\lambda}. 
\end{align*}
There exists a sufficiently large $t(\epsilon, \omega)$ such that $C \frac{\lambda^{t-t_0(\epsilon, \omega) + 1}}{1-\lambda} \le \frac{\epsilon}{2}$. Thus, it holds that for this fixed sample path $\omega$, for any $\epsilon>0$, there exists $t(\epsilon, \omega)$ such that for all $t\ge t(\epsilon, \omega)$
\begin{align*}
\sum_{m=0}^{t-1} \lambda^m \normfirst{\frac{1}{t-m}\sum_{r=1}^{t-m} w_j(r)}  \le \epsilon, 
\end{align*}
proving \eqref{eq: epsilon convergence}. 
%

\section*{Proof of Lemma \ref{lm: convergence rate}}
Our proof uses the McDiarmid's inequality. 

We first bound the expectation of $R_j(\lambda, t)$. 
\begin{align*}
\expect{R_j(\lambda, t)} & = \sum_{m=0}^{t-1} \lambda^m \expect{\norm{\frac{1}{t-m}\sum_{r=1}^{t-m} w_j(r)}}\\
& \overset{(a)}{\le} \sum_{m=0}^{t-1} \lambda^m \sqrt{\expect{\norm{\frac{1}{t-m}\sum_{r=1}^{t-m} w_j(r)}^2}} 
\end{align*}
where equality (a) follows from Jensen's inequality. Recall that $w_j(r)$ for $r=1, \cdots, t-m$ are independent and $\expect{w_j(r)} = \bm{0}$ for each $r=1, \cdots, t-m$. Thus, for any $j\in \calV/\calA$, we have 
\begin{align*}
\expect{\norm{\frac{1}{t-m}\sum_{r=1}^{t-m} w_j(r)}^2} 
& = \frac{1}{t-m} \trace \pth{\Sigma_j}. 
\end{align*}
So we get 
\begin{align*}
&\expect{R_j(\lambda, t)}  \le \sqrt{\trace\pth{\Sigma_j}}\sum_{m=1}^{t-1} \lambda^m \frac{1}{\sqrt{t-m}}. 
\end{align*}

We choose $h$ as 
\begin{align*}
h(\sth{w_j(r)}_{r=1}^t) ~ \triangleq ~  \sum_{m=0}^{t-1} \lambda^m \norm{\frac{\sum_{r=1}^{t-m}w_j(r)}{t-m}}. 
\end{align*}
It can be shown that we can choose $c_r$ to be 
\begin{align*}
c_r ~ = ~ C\sum_{m=0}^{t-r} \lambda^m \frac{1}{t-m}, ~~ \forall r=1, \cdots, t. 
\end{align*}
Let $m_0 = \frac{\log \frac{\lambda t}{2}}{\log \frac{1}{\lambda}}$. It is easy to see that $m_0\le \frac{t}{2}$ unless $t$ is extremely small. For simplicity, assume that $\frac{\log \frac{\lambda t}{2}}{\log \frac{1}{\lambda}}$ is an integer. 
So we have 
\begin{align*}
c_1 
& = C\pth{\sum_{m=0}^{m_0} \lambda^m \frac{1}{t-m} + \sum_{m_0+1}^{t-1} \lambda^m \frac{1}{t-m}} \le \frac{4C}{(1-\lambda)t}. 
\end{align*}
It is easy to see that $c_r \le c_1$ for all $r=1, \cdots, t$. So we have 
\begin{align*}
\sum_{r=1}^t c_r^2  \le t c_1^2  \le \pth{\frac{4C}{1-\lambda}}^2 \frac{1}{t}. 
\end{align*}

By McDiarmid's Inequality we have 
\begin{align*}
&\prob{R_j(\lambda, t) \ge \sqrt{\trace(\Sigma_j)}\sum_{m=1}^{t-1} \lambda^m \frac{1}{\sqrt{t-m}} + \epsilon} \\
&\le \exp \pth{\frac{-2\epsilon^2}{\sum_{r=1}^t c_r^2}} \le \exp \pth{\frac{-\epsilon^2(1-\lambda)^2t}{8C^2}}.  
\end{align*}

\section*{Proof of Theorem \ref{thm: complete consensus alt}}

%
For each $t$, $x_i(t)$ can be uniquely rewritten as 
\begin{align*}
x_i(t) = \theta^* + \sum_{k=1}^d \alpha_i^k (t) e_k,
\end{align*}
where $\alpha_i^k(t), k =1, \cdots, d$ is a linear coefficients. 
At time $t$, for each $k=1, \cdots, d$, it holds that 
\begin{align*}
\alpha_i^k(t) 
& = \frac{1}{\abth{\calR_i^k(t)}} \sum_{j\in \calR_i^k(t)} \iprod{m_{ji}(t)}{e_k}- \iprod{\theta^*}{e_k}. 
\end{align*}
%
It follows from Lemma \ref{lm: weight upper bound} that 
\begin{align}
\label{eq: k error}
& \alpha_i^k(t)   = \sum_{j\in \calV/\calA} \beta_{ij}^k(t)  \iprod{z_j(t)}{e_k}  - \iprod{\theta^* }{ e_k}. 
\end{align}
Recall from \eqref{eq: empirical} and \eqref{eq: empirical gradient}, 
for each $k=1, \cdots, d$, we have 
\begin{align*}
\iprod{z_i(t)}{e_k} & = \iprod{\theta^* }{ e_k}+ \iprod{H_i^{\top}\frac{1}{t}\sum_{r=1}^t w_i(r)}{~ e_k} \\
 & +  \iprod{\sum_{{k^{\prime}}=1}^d \alpha_i^{k^{\prime}} (t-1) \pth{{\bf I} - H_i^{\top} H_i} e_{k^{\prime}}}{~ e_k}. 
\end{align*}
Thus, \eqref{eq: k error} becomes 
\begin{align*}
&\alpha_i^k(t)  = \sum_{j\in \calV/\calA} \beta_{ij}^k(t)\iprod{H_j^{\top}\frac{1}{t}\sum_{r=1}^t w_j(r) }{~ e_k}\\
& +\sum_{j\in \calV/\calA} \beta_{ij}^k(t)\iprod{\sum_{k^{\prime}=1}^d \alpha_j^{k^{\prime}}(t-1) \pth{{\bf I} - H_j^{\top} H_j} e_{k^{\prime}}}{e_k}. 
\end{align*}
By Lemma \ref{lm: weight upper bound}, we have 
\begin{align*}
&\abth{\alpha_i^k(t)}  
\le \frac{\sum_{j\in \calV/\calA} \abth{\iprod{H_j^{\top}\frac{1}{t}\sum_{r=1}^t w_j(r) }{~ e_k}}}{\phi -b} \\
& + \frac{\sum_{j\in \calV/\calA}\abth{\iprod{\sum_{k^{\prime}=1}^d \alpha_j^{k^{\prime}}(t-1) \pth{{\bf I} - H_j^{\top} H_j} e_{k^{\prime}}}{e_k}}}{\phi -b}. 
\end{align*}
For the second term, we have 
\begin{align*}
&\abth{\iprod{\sum_{k^{\prime}=1}^d \alpha_j^{k^{\prime}}(t-1) \pth{{\bf I} - H_j^{\top} H_j} e_{k^{\prime}}}{e_k}} \\
& \le \pth{\max_{j\in \calV/\calA} \max_{1\le k^{\prime} \le d}  \abth{ \alpha_j^{k^{\prime}}(t-1)}} \normfirst{e_k^{\top}\pth{{\bf I} - H_j^{\top} H_j}}\\
& = \pth{\max_{j\in \calV/\calA}  \linf{x_j(t-1) - \theta^*}}\normfirst{\pth{{\bf I} - H_j^{\top} H_j}e_k}, 
\end{align*}
where the last equality follows from the fact that $\pth{{\bf I} - H_j^{\top} H_j}$ is symmetric. 
For the first term, we have 
\begin{align*}
\max_{1\le k \le d} \abth{\iprod{H_j^{\top}\frac{1}{t}\sum_{r=1}^t w_j(r)}{e_k}}& \le \norm{H_j^{\top}\frac{1}{t}\sum_{r=1}^t w_j(r)}\\
& \le C_0 \norm{\frac{1}{t}\sum_{r=1}^t w_j(r)}. 
\end{align*}
By Assumption \ref{ass: alternative assumption}, we have 
\begin{align*}
&\max_{i\in \calV/\calA} \linf{x_i(t) - \theta^*} \\
& \le  \rho \max_{i\in \calV/\calA} \linf{x_i(t-1) - \theta^*} + \max_{i\in \calV/\calA}C_0\norm{\frac{1}{t}\sum_{r=1}^t w_i(r)}\\
& \le \rho \max_{i\in \calV/\calA} \linf{x_i(t-1) - \theta^*} + \sum_{i\in \calV/\calA}C_0\norm{\frac{1}{t}\sum_{r=1}^t w_i(r)}\\
& \le  \rho^{t} \max_{i\in \calV/\calA} \linf{x_i(0) - \theta^*} + C_0\sum_{j\in \calV/\calA}R_j(\rho, t). 
\end{align*}
By Lemmas \ref{lm: noise bound} and \ref{lm: convergence rate} with $\lambda= \rho$, we complete the proof. 

\section*{Proof of Theorem \ref{thm: general graphs local communication}}

We first show that the evolution of $\|x_i(t) - \theta^*\|_{\infty}$ -- the $\ell_{\infty}$ norm of the estimation errors -- for all $i\in \calV/\calA$ collectively have a matrix representation. Then we bound the convergence rate of the obtained matrix product. \\

Similar to the proof of Theorem \ref{thm: complete consensus alt}, for any $i\in \calV/\calA$ and any $k$, we have 
\begin{align*}
&|\alpha_i^k(t)| \le 
\abth{\sum_{j\in \calN_i/\calA} \beta_{ij}^{k}(t)\iprod{H_j^{\top}\frac{1}{t}\sum_{r=1}^t w_j(r) }{~ e_{k}}} \\
& + \sum_{j\in \calN_i/\calA} \beta_{ij}^k(t) \abth{\iprod{\pth{{\bf I} - H_j^{\top} H_j} I
\pth{\sum_{k^{\prime}=1}^d \alpha_{j}^{k^{\prime}}(t-1)e_{k^{\prime}}}}{e_k}}. 
\end{align*}
For the second term, we have 
\begin{align*}
&\sum_{j\in \calN_i/\calA} \beta_{ij}^k(t) \abth{\iprod{\pth{{\bf I} - H_j^{\top} H_j} 
\pth{\sum_{k^{\prime}=1}^d \alpha_{j}^{k^{\prime}}(t-1)e_{k^{\prime}}}}{e_k}} \\
& \le \sum_{j\in \calN_i/\calA} \beta_{ij}^k(t) \normfirst{\pth{{\bf I} - H_j^{\top} H_j} e_k} \linf{x_j(t-1) -\theta^*}. 
\end{align*}
For the first term, we have 
\begin{align*}
&\abth{\sum_{j\in \calN_i/\calA} \beta_{ij}^{k}(t)\iprod{H_j^{\top}\frac{1}{t}\sum_{r=1}^t w_j(r) }{~ e_{k}}} \\
& \le C_0 \max_{j\in \calV/\calA} \norm{\frac{1}{t}\sum_{r=1}^t w_j(r)}. 
\end{align*}
Thus, we get 
\begin{align*}
&\linf{x_i(t) - \theta^*} = \max_{1\le k \le d} |\alpha_i^k(t)| \\
& \le \max_{1\le k \le d} \sum_{j\in \calN_i/\calA} \beta_{ij}^k(t) \normfirst{\pth{{\bf I} - H_j^{\top} H_j} e_k} \linf{x_j(t-1) -\theta^*} \\
& +\quad C_0 \max_{j\in \calV/\calA} \norm{\frac{1}{t}\sum_{r=1}^t w_j(r)}. 
\end{align*}
Let $\bm{E}(t) \in \reals^{\phi}$ be the vector that stacks the $\ell_{\infty}$ norm of the errors $x_i(t) - \theta^*$ for all $i\in \calV/\calA$. For each $i\in \calV/\calA$, define matrix $M(t)$ as follows: 
\begin{align*}
M_{i,j} (t) = \beta_{i,j}^{k_i^*(t)}  \normfirst{\pth{{\bf I} - H_j^{\top} H_j} e_{k_i^*(t)}},
\end{align*}
where $k_i^*(t)$ is an arbitrary maximizer of  
$$\sum_{j\in \calN_i/\calA} \beta_{ij}^k(t) \normfirst{\pth{{\bf I} - H_j^{\top} H_j} e_k} \linf{x_j(t-1) -\theta^*}$$
over $k=1, \cdots, d$. 
%
%
With this rewriting, we have 
\begin{align*}
&\bm{E}(t) \le M(t) \bm{E}(t-1) + C_0 \max_{j\in \calV/\calA} \norm{\frac{1}{t}\sum_{r=1}^t w_j(r)} \ones \\
& \le \pth{\prod_{r=1}^{t} M(r)} \bm{E}(0) + C_0 \max_{j\in \calV/\calA} \norm{\frac{1}{t}\sum_{r=1}^t w_j(r)} \ones\\
&+C_0 \sum_{m=1}^{t-1} \max_{j\in \calV/\calA} \norm{\frac{\sum_{r=1}^{t-m} w_j(r)}{t-m} }\pth{\prod_{r=t-m+1}^{t} M(r)} \bm{1}. 
\end{align*}
Note that $M(t)$ is random, and its realization is determined by both the noises in the good agents' local observations and the Byzantine agents' adversarial behaviors. Nevertheless, this does not complicate our analysis because our analysis works for every realization of $M(t)$.  Henceforth, with a little abuse of notation, we use $M(t)$ to denote both the random matrix and its realization. 

By Lemma \ref{lm: weight upper bound general} and Assumption \ref{ass: general graphs}, we know that for every $t$, the matrix $M(t)$ is a {\em strict} sub-stochastic matrix. In particular, under the assumptions in Theorem \ref{thm: general graphs local communication}, the following claim is true. 
\begin{claim}
\label{claim: consecutive product}
For any $t_0$ and for any sequence of realization of the matrices $M(t)$ for $t=t_0+1, \cdots, t_0+\xi \phi$, the following holds  
\[
\pth{\prod_{t=t_0+1}^{t_0+\xi \phi} M(t)}\bm{1} \le \gamma \bm{1}, ~  \text{where} ~ \gamma : = 1- \frac{1-\rho_0}{\pth{2(\phi-b)}^{\xi\phi}}.
\] 
\end{claim}
For ease of exposition, the proof of Claim \ref{claim: consecutive product} is deferred to the end of this paper. 

With Claim \ref{claim: consecutive product}, for any fixed $t_0$ and for sufficiently large $t-t_0$, we have 
\begin{align*}
&\pth{\prod_{r=t_0+1}^{t} M(r)} \bm{1} \\
&= \pth{\prod^{t}_{r=t_0+\xi\phi+1}  M(r)}\pth{\prod_{r=t_0+1}^{t_0+\xi\phi} M(r)}\bm{1} \\
& \le \gamma \pth{\prod^{t}_{r=t_0+\xi\phi+1} M(r)} \bm{1}\\
& \le \gamma^{\lfloor \frac{t-t_0}{\xi\phi}\rfloor} \pth{\prod^{t}_{r=\lfloor \frac{t-t_0}{\xi\phi}\rfloor \xi\phi +1}M(r)} \bm{1} \\
&\le \gamma^{\lfloor \frac{t-t_0}{\xi\phi}\rfloor} \bm{1}. 
\end{align*}
Thus, 
\begin{align*}
\pth{\prod_{r=1}^{t} M(r)} \bm{E}(0) & \le \pth{\prod_{r=1}^{t} M(r)} \max_{i\in \calV/\calA}\linf{x_i(0)-\theta^*} \bm{1}\\
& \le \max_{i\in \calV/\calA}\linf{x_i(0)-\theta^*}  \gamma^{\lfloor \frac{t}{\xi\phi}\rfloor} \bm{1}.  
\end{align*}
In addition, 
\begin{align*}
&\sum_{m=0}^{t-1} \max_{j\in \calV/\calA} \norm{\frac{1}{t-m} \sum_{r=1}^{t-m} w_j(r)}\pth{\prod_{r=t-m+1}^{t} M(r)} \bm{1}\\
& \le \sum_{m=0}^{t-1} \max_{j\in \calV/\calA} \norm{\frac{1}{t-m} \sum_{r=1}^{t-m} w_j(r)}  \gamma^{\lfloor \frac{m}{\xi\phi}\rfloor}\bm{1}\\
& \le \sum_{m=0}^{t-1} \sum_{j\in \calV/\calA} \norm{\frac{1}{t-m} \sum_{r=1}^{t-m} w_j(r)}  \gamma^{\lfloor \frac{m}{\xi\phi}\rfloor}\bm{1}. 
\end{align*}

For ease of exposition, we assume that $\lfloor \frac{m}{\xi\phi}\rfloor$ is an integer for any $m$. Note that this simplification does not affect the order of convergence. 
\begin{align*}
\bm{E}(t) & \le \pth{\max_{i\in \calV/\calA}\linf{x_i(0)-\theta^*}}  \gamma^{\frac{t}{\xi\phi}} \bm{1} \\
& + C_0 \sum_{j\in \calV/\calA} \sum_{m=0}^{t-1} \norm{\frac{1}{t-m} \sum_{r=1}^{t-m} w_j(r)}  \gamma^{\frac{m}{\xi\phi}}\bm{1}\\
& \le \pth{\max_{i\in \calV/\calA}\linf{x_i(0)-\theta^*}}  \gamma^{ \frac{t}{\xi\phi}} \bm{1} \\
& + C_0\sum_{j\in \calV/\calA} R_j(\gamma^{ \frac{1}{\xi\phi}}, t). 
\end{align*}

Applying Lemma \ref{lm: noise bound} with $\lambda = \gamma^{ \frac{1}{\xi\phi}}$, we have 
\begin{align*}
0 \le \lim_{t\diverge}\bm{E}(t) \le 0 + 0 + 0 = 0, ~~ \text{almost surely}. 
\end{align*}

In addition, by applying Lemma \ref{lm: convergence rate} with $\lambda = \gamma^{ \frac{1}{\xi\phi}}$, we complete the proof.

\section*{Proof of Claim \ref{claim: consecutive product}}
Recall that $M(t)$ (for each $t\ge 1$) is defined as 
\begin{align*}
M_{i,j} (t) = \beta_{i,j}^{k_i^*(t)}  \normfirst{\pth{{\bf I} - H_j^{\top} H_j} e_{k_i^*(t)}},
\end{align*}
where $k_i^*(t)$ is an arbitrary maximizer of  
$$\sum_{j\in \calN_i/\calA} \beta_{ij}^k(t) \normfirst{\pth{{\bf I} - H_j^{\top} H_j} e_k} \linf{x_j(t-1) -\theta^*}$$
over $k=1, \cdots, d$. 

For any sequence of realization of the matrices $M(t)$ for $t=t_0+1, \cdots, t_0+\xi \phi$, we construct a sequence of auxiliary {\em stochastic} matrices, denoted by $\tilde{M}(t)$, 
as follows: 
\begin{align*}
\tilde{M}_{ij}(t) : =  \beta_{ij}^{k^*_i(t)}, ~~ \forall i, j \in \calV/\calA. 
\end{align*}
By Lemma \ref{lm: weight upper bound general}, $\tilde{M}(t)$ is row-stochastic for $t=t_0+1, \cdots, t_0+\xi \phi$. 
By Definition \ref{reduced graph} and Lemma \ref{lm: weight upper bound general}, for each $t$ there exists a reduced graph in $\calH$ such that 
\begin{align*}
\tilde{M}(t) \ge \frac{1}{2\pth{\phi-b}} H(t),
\end{align*}
where $H(t)$ is the adjacency matrix of the corresponding reduced graph. For ease of exposition, with a little abuse of notation, we use $H(t)$ to denote both the adjacency matrix and the reduced graph. \footnote{Its meaning should be clear from the context.} We refer to $H(t)$ as the {\em shadow graph} at time $t$. 

Since the matrix product $\prod_{t=t_0+1}^{t_0+\xi \phi} M(t)$ consists of $\xi\phi$ shadow graphs and $|\calH| = \xi$, there exists at least one reduced graph in $\calH$ that appears at least $\phi$ times in the sequence of shadow graphs. Let $\tilde{H}$ be one such reduced graph. Without loss of generality, let $i_0$ be the node in the unique source component of $\tilde{H}$ such that 
\begin{align*}
&\abth{(\calN_{i_0}\cup \{i_0\} /\calA)\cap \sth{j:\normfirst{\pth{{\bf I} -H_j^{\top} H_j} e_{k}} <1}}\\
&\ge b+1. 
\end{align*}
Since $i_0$ in the unique source component of $\tilde{H}$, it follows that node $i_0$ can reach every other good agents within $\phi-1$ hops using the edges in $\tilde{H}$ only.

For any given realization of $M(t_0+1), \cdots, M(t_0+\xi\phi)$, let $\tau_1, \cdots, \tau_{\phi}$ be the first $\phi$ time indices at which $\tilde{H}$ is the shadow graph. In addition, let 
\[
\Delta_j : = \tau_j - \tau_{j-1}, ~~ ~~ \forall ~ j= 2, \cdots, \phi.   
\]

For ease of exposition, in the reminder of this proof, we assume $t_0=0$. The proof can be easily generalized to arbitrary $t_0$. 
Let 
\[
\eta (t) : =   \pth{\prod_{r=1}^{t} M(r)}\bm{1}, ~~~ \forall t. 
\]
Note that $\eta(t) \le \bm{1}$ as $M(r)$ is sub-stochastic for all $r$.

To show Claim \ref{claim: consecutive product}, it is enough to show the following three claims. 
\begin{itemize}
\item [(A)] For any $j= 1, \cdots, \phi$, 
\[
\eta_{i_0} (\tau_j) \le 1- \frac{1- \rho_0}{2(\phi-b)};
\]  
\item [(B)] If $i$ is an outgoing neighbor of $i_0$ in the shadow graph $\tilde{H}$, i.e., $\tilde{H}_{ii_0}=1$, 
then for any $j= 2, \cdots, \phi$, 
\[
\eta_{i} (\tau_j) \le 1 - \frac{1-\rho_0}{\pth{2(\phi-b)}^2}.  
\] 
\item [(C)] For any $j= 3, \cdots, \phi$, if $i_0$ can reach node $i$ in the shadow graph $\tilde{H}$ with $h$ hops, where $2\le h\le j-1$, then 
\[
\eta_{i} (\tau_j) 
\le 1- \frac{1-\rho_0}{\pth{2(\phi-b)}^{2+\sum_{j^{\prime}=j+2-h}^{j} \Delta_{j^{\prime}}}}. 
\]
\end{itemize}

Suppose Claims (A), (B), and (C) hold. Recall that $i_0$ is in the unique source component of $\tilde{H}$. At time $\tau_{\phi}$, at all $i\in \calV/\calA$, it holds that 
\begin{align*}
\eta_{i} (\tau_{\phi}) &\le 1 - \frac{1-\rho_0}{\pth{2(\phi-b)}^{2+\sum_{j^{\prime}=3}^{\phi} \Delta_{j^{\prime}}}}\\
& \le 1 - \frac{1-\rho_0}{\pth{2(\phi-b)}^{\xi\phi}}. 
\end{align*}
Therefore, we conclude that 
\begin{align*}
\eta (\xi\phi) 
& = \pth{\prod_{r=\tau_{\phi}+1}^{\xi\phi} M(r)} \eta (\tau_{\phi})\\
& \le \pth{1 - \frac{1-\rho_0}{\pth{2(\phi-b)}^{\xi\phi}}} \pth{\prod_{r=\tau_{\phi}+1}^{\xi\phi} M(r)} \bm{1}\\
& \le \pth{1 - \frac{1-\rho_0}{\pth{2(\phi-b)}^{\xi\phi}}} \bm{1},
\end{align*}
proving Claim \ref{claim: consecutive product}. 

\vskip \baselineskip 

In the remainder of the proof, we prove Claims (A), (B), and (C), individually.

\paragraph{We first show (A)} 
For any $j= 1, \cdots, \phi$, we have 
\begin{align*}
\eta (\tau_j) 
& \le M(\tau_j)  \bm{1}. 
\end{align*}
Thus 
\begin{align*}
&\eta_{i_0} (\tau_j)  \le  \sum_{i \in \calV/\calA} M_{i_0 i}(\tau_j)  \\
& = \sum_{i\in \calV/\calA}  \beta_{i_0i}^{k_{i_0}^*(\tau_j)}  \normfirst{\pth{{\bf I} - H_{i}^{\top} H_{i}} e_{k_{i_0}^*(\tau_j)}}\\
& \le \sum_{i \in \calV/\calA ~ \& ~ \normfirst{\pth{{\bf I} - H_{i}^{\top} H_{i}} e_{k_{i_0}^*(\tau_j)}}<1} \beta_{i_0 i}^{k_{i_0}^*(\tau_j)} \rho_0\\
& \quad + \sum_{i \in \calV/\calA ~ \& ~ \normfirst{\pth{{\bf I} - H_{i}^{\top} H_{i}} e_{k_{i_0}^*(\tau_j)}}=1} \beta_{{i_0} i}^{k_{i_0}^*(\tau_j)}. 
\end{align*}
By Lemma \ref{lm: weight upper bound general}, Assumption \ref{ass: general graphs}, and the choice of $i_0$, we know that 
\begin{align*}
&\sum_{i \in \calV/\calA ~ \& ~ \normfirst{\pth{{\bf I} - H_{i}^{\top} H_{i}} e_{k_{i_0}^*(\tau_j)}}<1} \beta_{i_0 i}^{k_{i_0}^*(\tau_j)} \\
&\ge \frac{1}{2(|\calN_{i_0} \cup\{{i_0}\}/\calA| -b)} \\
& \ge \frac{1}{2(\phi-b)}. 
\end{align*}
Thus, we have 
\begin{align*}
\eta_{i_0} (\tau_j) 
& \le 1- \frac{1-\rho_0}{2(\phi-b)}. 
\end{align*}

%

\paragraph{Next we show (B)}
For any $j= 2, \cdots, \nu$, 
\begin{align*}
\eta (\tau_j)  =  M(\tau_j) \eta (\tau_j-1) = \sum_{i^{\prime}\in \calV/\calA} M_{i i^{\prime}}(\tau_j) \eta_{i^{\prime}} (\tau_j-1). 
\end{align*}
Recall that 
\[
M_{ii_0}(\tau_j) = \beta_{ii_0}^{k_i^*(\tau_j)} \normfirst{\pth{\bm{I} - H_{i_0}^{\top} H_{i_0}} k_i^*(\tau_j)}.
\]
We consider two cases: 
\begin{itemize}
\item[(1)] $\normfirst{\pth{\bm{I} - H_{i_0}^{\top} H_{i_0}} k_i^*(\tau_j)} <1$; 
\item[(2)] $\normfirst{\pth{\bm{I} - H_{i_0}^{\top} H_{i_0}} k_i^*(\tau_j)} =1$. 
\end{itemize}

Suppose that $\normfirst{\pth{\bm{I} - H_{i_0}^{\top} H_{i_0}} k_i^*(\tau_j)} <1$. Since $\tilde{H}_{ii_0} =1$, it follows that 
\[
\tilde{M}_{ii_0} (\tau_j) =  \beta_{ii_0}^{k_i^*(\tau_j)} \ge \frac{1}{2(\phi-b)}. 
\]
Thus, we have 
\begin{align*}
\eta_i (\tau_j)  
& \le  M_{i i_0}(\tau_j) + \sum_{i^{\prime}\in \calV/\calA \& i^{\prime} = i_0} M_{i i^{\prime}}(\tau_j) \\
& \le \beta_{ii_0}^{k_i^*(\tau_j)} \rho_0 + \sum_{i^{\prime}\in \calV/\calA \& i^{\prime} = i_0} \beta_{ii^{\prime}}^{k_i^*(\tau_j)} \\
& = 1 -  \beta_{ii}^{k_i^*(\tau_j)} (1-\rho_0) \\
& \le 1- \frac{1-\rho_0}{2(\phi-b)}.  
\end{align*}

Suppose that $\normfirst{\pth{\bm{I} - H_{i_0}^{\top} H_{i_0}} k_i^*(\tau_j)} =1$. In this case
\[
M_{ii_0}(\tau_j) = \tilde{M}_{ii_0}(\tau_j) \ge \frac{1}{2(\phi-b)}. 
\]
Thus, we have  
\begin{align*}
\eta_i (\tau_j) & = M_{i i_0}(\tau_j) \eta_{i_0} (\tau_j-1) \\
& \quad+ \sum_{i^{\prime}\in \calV/\calA, \& i^{\prime} = i_0} M_{i i^{\prime}}(\tau_j) \eta_{i^{\prime}} (\tau_j-1)\\
& \le M_{i i_0}(\tau_j) \pth{1- \frac{1-\rho_0}{2(\phi-b)}} \\
& \quad+ \sum_{i^{\prime}\in \calV/\calA \& i^{\prime} = i_0} M_{i i^{\prime}}(\tau_j) \\
& \le  \sum_{i^{\prime}\in \calV/\calA} M_{i, i^{\prime}}(\tau_j) - \frac{1-\rho_0}{2(\phi-b)} M_{ii_0}(\tau_j)\\
& \le 1- \frac{1-\rho_0}{(2(\phi-b))^2}. 
\end{align*}

\paragraph{Finally we show (C)}

We prove this by induction. 

\noindent {\bf Base case: $j=3$} 

Let $i$ be a $2$--th order neighbor of node $i_0$ in the shadow graph $\tilde{H}$; there exists a directed path of length $2$ such that $i_0 \to i_1\to i$ in $\tilde{H}$. 

If $\normfirst{\pth{\bm{I} - H_{i_1}^{\top} H_{i_1}} k_i^*(\tau_3)} <1$, similar to the proof of Claim (B), we have that 
\begin{align*}
\eta_i (\tau_3) & \le 1- \frac{1-\rho_0}{2(\phi-b)}.  
\end{align*}
Now suppose $\normfirst{\pth{\bm{I} - H_{i_1}^{\top} H_{i_1}} k_i^*(\tau_3)} =1$. 

If there exists $r$ where $\tau_2+1 \le r \le \tau_3-1$ such that 
\[
\normfirst{\pth{\bm{I} - H_{i_1}^{\top} H_{i_1}} k_{i_1}^*(r)} <1,  
\]
i.e., $M_{i_1i_1} (r) < \tilde{M}_{i_1i_1} (r)$. Let $r^*$ be the latest time index. Note that $\beta_{ii}^k (t) \ge \frac{1}{2(\phi-b)}$ for any $i\in \calV/\calA$, $t$ and $k$.  
We have 
\begin{align*}
\eta_{i_1} (r^*) 
& \le \sum_{i^{\prime}\in \calV/\calA} M_{i_1i^{\prime}}(r^*) \le 1 - \frac{1-\rho_0}{2(\phi-b)}. 
\end{align*}
In addition, by the choice of $r^*$, we have 
\[
\qth{\prod_{r=r^*+1}^{\tau_3-1} M(r)}_{i_1 i_1} \ge \frac{1}{\pth{2(\phi-b)}^{\tau_3-r^*-1}}.  
\]
So we get 
\begin{align*}
\eta_{i_1} (\tau_3-1) & 
= \qth{\prod_{r=r^*+1}^{\tau_3-1} M(r)}_{i_1 i_1} \eta_{i_1} (r^*) \\
& \quad +  \sum_{i^{\prime}\in \calV/\calA}\qth{\prod_{r=r^*+1}^{\tau_3-1} M(r)}_{i_1 i^{\prime}} \eta_{i^{\prime}} (r^*)\\
& \le 1 - \frac{1-\rho_0}{\pth{2(\phi-b)}^{\tau_3-r^*}}. 
\end{align*}

As $\normfirst{\pth{\bm{I} - H_{i_1}^{\top} H_{i_1}} k_i^*(\tau_3)} =1$ and $\beta_{ii_0}^{i^*(\tau_3)} \ge \frac{1}{2(\phi-b)}$, we get that 
\[
\eta_i(\tau_3) \le 1 - \frac{1-\rho_0}{\pth{2(\phi-b)}^{\tau_3-r^*+1}} \le 1 - \frac{1-\rho_0}{\pth{2(\phi-b)}^{\Delta_3}}. 
\]

To finish the proof of the base case, it remains to consider the case that 
\[
\normfirst{\pth{\bm{I} - H_{i_1}^{\top} H_{i_1}} k_{i_1}^*(r)} =1,  
\]
i.e., $M_{i_1 i_1}(r) = \tilde{M}_{i_1 i_1}(r)$ for all $r$ such that $\tau_2+1 \le r \le \tau_3-1$. Thus, we get 
\[
\qth{\prod_{r=\tau_2+1}^{\tau_3-1} M(r)}_{i_1 i_1} \ge \frac{1}{\pth{2(\phi-b)}^{\Delta_3-1}}.  
\]
So 
\begin{align*}
\eta_{i_1}(\tau_3-1) & = \sum_{i^{\prime} \in \calV/\calA} \qth{\prod_{r=\tau_2+1}^{\tau_3-1} M(r)}_{i_1, i^{\prime}} \eta_{i^{\prime}}(\tau_2)\\
& \le 1 - \qth{\prod_{r=\tau_2+1}^{\tau_3-1} M(r)}_{i_1, i_1} \frac{1}{\pth{2(\phi-b)}^{2}}\\
& \le 1 - \frac{1}{\pth{2(\phi-b)}^{\Delta_3+1}},
\end{align*}
and 
\begin{align*}
\eta_{i}(\tau_3) \le  1 - \frac{1}{\pth{2(\phi-b)}^{\Delta_3+2}}. 
\end{align*}

\noindent {\bf Induction step:}
Suppose the following holds for any $j= 3, \cdots, \phi-1$: 
\[
\eta_i(\tau_j) \le 1- \frac{\rho_0}{\pth{2(\phi-b)}^{2+\sum_{j^{\prime}=j+2-h}^{j} \Delta_{j^{\prime}}}}
\]
for all the $h$--th order neighbor of node $i_0$ in the shadow graph $\tilde{H}$, where $h=2, \cdots, j-1$.

\vskip \baselineskip 

\noindent {\bf Inductive step:}

The proof of the inductive step is similar to the proof of the base case, thus is omitted.

\bibliography{WTABib,shahin}

\begin{thebibliography}{10}
\providecommand{\url}[1]{#1}
\csname url@samestyle\endcsname
\providecommand{\newblock}{\relax}
\providecommand{\bibinfo}[2]{#2}
\providecommand{\BIBentrySTDinterwordspacing}{\spaceskip=0pt\relax}
\providecommand{\BIBentryALTinterwordstretchfactor}{4}
\providecommand{\BIBentryALTinterwordspacing}{\spaceskip=\fontdimen2\font plus
\BIBentryALTinterwordstretchfactor\fontdimen3\font minus
  \fontdimen4\font\relax}
\providecommand{\BIBforeignlanguage}[2]{{%
\expandafter\ifx\csname l@#1\endcsname\relax
\typeout{** WARNING: IEEEtran.bst: No hyphenation pattern has been}%
\typeout{** loaded for the language `#1'. Using the pattern for}%
\typeout{** the default language instead.}%
\else
\language=\csname l@#1\endcsname
\fi
#2}}
\providecommand{\BIBdecl}{\relax}
\BIBdecl

\bibitem{speranzon2006distributed}
A.~Speranzon, C.~Fischione, and K.~H. Johansson, ``Distributed and
  collaborative estimation over wireless sensor networks,'' in \emph{IEEE
  Conference on Decision and Control (CDC)}, 2006, pp. 1025--1030.

\bibitem{xie2012fully}
L.~Xie, D.-H. Choi, S.~Kar, and H.~V. Poor, ``Fully distributed state
  estimation for wide-area monitoring systems,'' \emph{IEEE Transactions on
  Smart Grid}, vol.~3, no.~3, pp. 1154--1169, 2012.

\bibitem{sinopoli2003distributed}
B.~Sinopoli, C.~Sharp, L.~Schenato, S.~Schaffert, and S.~S. Sastry,
  ``Distributed control applications within sensor networks,''
  \emph{Proceedings of the IEEE}, vol.~91, no.~8, pp. 1235--1246, 2003.

\bibitem{olfati2007distributed}
R.~Olfati-Saber, ``Distributed kalman filtering for sensor networks,'' in
  \emph{IEEE Conference on Decision and Control (CDC)}, 2007, pp. 5492--5498.

\bibitem{kar2012distributed}
S.~Kar, J.~M. Moura, and K.~Ramanan, ``Distributed parameter estimation in
  sensor networks: Nonlinear observation models and imperfect communication,''
  \emph{IEEE Transactions on Information Theory}, vol.~58, no.~6, pp.
  3575--3605, 2012.

\bibitem{bullo2009distributed}
F.~Bullo, J.~Cortes, and S.~Martinez, \emph{Distributed control of robotic
  networks: a mathematical approach to motion coordination algorithms}.\hskip
  1em plus 0.5em minus 0.4em\relax Princeton University Press, 2009, vol.~27.

\bibitem{chen2018internet}
Y.~Chen, S.~Kar, and J.~M. Moura, ``The internet of things: Secure distributed
  inference,'' \emph{IEEE Signal Processing Magazine}, vol.~35, no.~5, pp.
  64--75, 2018.

\bibitem{stankovic2011decentralized}
S.~S. Stankovic, M.~S. Stankovic, and D.~M. Stipanovic, ``Decentralized
  parameter estimation by consensus based stochastic approximation,''
  \emph{IEEE Transactions on Automatic Control}, vol.~56, no.~3, pp. 531--543,
  2011.

\bibitem{chen2017distributed}
Y.~Chen, L.~Su, and J.~Xu, ``Distributed statistical machine learning in
  adversarial settings: Byzantine gradient descent,'' \emph{Proceedings of the
  ACM on Measurement and Analysis of Computing Systems}, vol.~1, no.~2, p.~44,
  2017.

\bibitem{kar2009distributed}
S.~Kar and J.~M. Moura, ``Distributed consensus algorithms in sensor networks
  with imperfect communication: Link failures and channel noise,'' \emph{IEEE
  Transactions on Signal Processing}, vol.~57, no.~1, pp. 355--369, 2009.

\bibitem{Lynch:1996:DA:2821576}
N.~A. Lynch, \emph{Distributed Algorithms}.\hskip 1em plus 0.5em minus
  0.4em\relax San Francisco, CA, USA: Morgan Kaufmann Publishers Inc., 1996.

\bibitem{pease1980reaching}
M.~Pease, R.~Shostak, and L.~Lamport, ``Reaching agreement in the presence of
  faults,'' \emph{Journal of the ACM (JACM)}, vol.~27, no.~2, pp. 228--234,
  1980.

\bibitem{lamport1982byzantine}
L.~Lamport, R.~Shostak, and M.~Pease, ``The byzantine generals problem,''
  \emph{ACM Transactions on Programming Languages and Systems (TOPLAS)},
  vol.~4, no.~3, pp. 382--401, 1982.

\bibitem{mendes2013multidimensional}
H.~Mendes and M.~Herlihy, ``Multidimensional approximate agreement in byzantine
  asynchronous systems,'' in \emph{Proceedings of the forty-fifth annual ACM
  symposium on Theory of computing}.\hskip 1em plus 0.5em minus 0.4em\relax
  ACM, 2013, pp. 391--400.

\bibitem{vaidya2013byzantine}
N.~H. Vaidya and V.~K. Garg, ``Byzantine vector consensus in complete graphs,''
  in \emph{Proceedings of the 2013 ACM symposium on Principles of distributed
  computing}.\hskip 1em plus 0.5em minus 0.4em\relax ACM, 2013, pp. 65--73.

\bibitem{kosut2011malicious}
O.~Kosut, L.~Jia, R.~J. Thomas, and L.~Tong, ``Malicious data attacks on the
  smart grid,'' \emph{IEEE Transactions on Smart Grid}, vol.~2, no.~4, pp.
  645--658, 2011.

\bibitem{kim2011strategic}
T.~T. Kim and H.~V. Poor, ``Strategic protection against data injection attacks
  on power grids,'' \emph{IEEE Transactions on Smart Grid}, vol.~2, no.~2, pp.
  326--333, 2011.

\bibitem{sou2013exact}
K.~C. Sou, H.~Sandberg, and K.~H. Johansson, ``On the exact solution to a smart
  grid cyber-security analysis problem,'' \emph{IEEE Transactions on Smart
  Grid}, vol.~4, no.~2, pp. 856--865, 2013.

\bibitem{pasqualetti2013attack}
F.~Pasqualetti, F.~D{\"o}rfler, and F.~Bullo, ``Attack detection and
  identification in cyber-physical systems,'' \emph{IEEE Transactions on
  Automatic Control}, vol.~58, no.~11, pp. 2715--2729, 2013.

\bibitem{fawzi2014secure}
H.~Fawzi, P.~Tabuada, and S.~Diggavi, ``Secure estimation and control for
  cyber-physical systems under adversarial attacks,'' \emph{IEEE Transactions
  on Automatic Control}, vol.~59, no.~6, pp. 1454--1467, 2014.

\bibitem{mattingley2010real}
J.~Mattingley and S.~Boyd, ``Real-time convex optimization in signal
  processing,'' \emph{IEEE Signal processing magazine}, vol.~27, no.~3, pp.
  50--61, 2010.

\bibitem{shoukry2016event}
Y.~Shoukry and P.~Tabuada, ``Event-triggered state observers for sparse sensor
  noise/attacks,'' \emph{IEEE Transactions on Automatic Control}, vol.~61,
  no.~8, pp. 2079--2091, 2016.

\bibitem{sundaram2011distributed}
S.~Sundaram and C.~N. Hadjicostis, ``Distributed function calculation via
  linear iterative strategies in the presence of malicious agents,'' \emph{IEEE
  Transactions on Automatic Control}, vol.~56, no.~7, pp. 1495--1508, 2011.

\bibitem{chen2018resilient}
Y.~Chen, S.~Kar, and J.~M. Moura, ``Resilient distributed estimation through
  adversary detection,'' \emph{IEEE Transactions on Signal Processing}, 2018.

\bibitem{chen2018attack}
------, ``Attack resilient distributed estimation: A consensus+ innovations
  approach,'' in \emph{2018 Annual American Control Conference (ACC)}, 2018,
  pp. 1015--1020.

\bibitem{mitra2018resilient}
A.~Mitra and S.~Sundaram, ``Resilient distributed state estimation for {LTI}
  systems,'' \emph{arXiv preprint arXiv:1802.09651}, 2018.

\bibitem{xu2018robust}
W.~Xu, Z.~Li, and Q.~Ling, ``Robust decentralized dynamic optimization at
  presence of malfunctioning agents,'' \emph{Signal Processing}, vol. 153, pp.
  24--33, 2018.

\bibitem{su2016non}
L.~Su and N.~H. Vaidya, ``Non-bayesian learning in the presence of byzantine
  agents,'' in \emph{International Symposium on Distributed Computing}.\hskip
  1em plus 0.5em minus 0.4em\relax Springer, 2016, pp. 414--427.

\bibitem{yang2017byrdie}
Z.~Yang and W.~U. Bajwa, ``Byrdie: Byzantine-resilient distributed coordinate
  descent for decentralized learning,'' \emph{arXiv preprint arXiv:1708.08155},
  2017.

\bibitem{su2016fault}
L.~Su and N.~H. Vaidya, ``Fault-tolerant multi-agent optimization: optimal
  iterative distributed algorithms,'' in \emph{Proceedings of the 2016 ACM
  Symposium on Principles of Distributed Computing}.\hskip 1em plus 0.5em minus
  0.4em\relax ACM, 2016, pp. 425--434.

\bibitem{vaidya2012iterative}
N.~H. Vaidya, L.~Tseng, and G.~Liang, ``Iterative approximate byzantine
  consensus in arbitrary directed graphs,'' in \emph{Proceedings of the 2012
  ACM symposium on Principles of distributed computing}.\hskip 1em plus 0.5em
  minus 0.4em\relax ACM, 2012, pp. 365--374.

\bibitem{vaidya2012matrix}
N.~Vaidya, ``Matrix representation of iterative approximate byzantine consensus
  in directed graphs,'' \emph{arXiv preprint arXiv:1203.1888}, 2012.

\end{thebibliography}
\bibliographystyle{IEEEtran}

\end{document}